\documentclass[english, 12pt]{amsart}
\usepackage[T1]{fontenc}
\usepackage[latin9]{inputenc}
\usepackage{amsthm}

\makeatletter
\numberwithin{equation}{section}
\numberwithin{figure}{section}
\theoremstyle{plain}
\newtheorem{thm}{Theorem}
  \theoremstyle{definition}
  \newtheorem{defn}[thm]{Definition}
  \theoremstyle{plain}
  \newtheorem{lem}[thm]{Lemma}
  \theoremstyle{plain}
  \newtheorem{prop}[thm]{Proposition}
  \theoremstyle{plain}
  \newtheorem{cor}[thm]{Corollary}

\makeatother

\usepackage{babel}

\begin{document}

\title{$H_{2}$-reducible Hadamard matrices of order 6}

\author{Bengt R. Karlsson}

\address{Uppsala University, Dept of Physics and Astronomy, Box 516, SE-751
20, Uppsala, Sweden}

\email{bengt.karlsson@physics.uu.se}
\begin{abstract}
Complex Hadamard matrices $H$ of order 6 are characterized in a novel
manner, according to the presence/absence of order 2 Hadamard submatrices.
It is shown that if there exists one such submatrix, $H$ is equivalent
to a Hadamard matrix where all the nine submatrices are Hadamard.
The ensuing subset of $H_{2}$-reducible complex Hadamard matrices
is more general than might be thought, and, significantly, includes
all the up till now described (one- and two-parameter) families of
order 6. A known, isolated matrix, and most numerically generated
matrices, fall outside the subset.
\end{abstract}
\maketitle

\section{Introduction}

Complex Hadamard matrices (for an overview, see \cite{Tadej guide,Tadej webguide})
have recently become a topic of interest, in part because of the correspondence
between such matrices and mutually unbiased bases, MUBs. Particular
attention has been given to two unsettled problems in six dimensions,
see for instance \cite{Bengtsson_08}. On the one hand, six is the
lowest order for which a complete characterization of the complex
Hadamard matrices is lacking, and, on the other hand, it is also the
lowest dimension for which a full understanding of the MUBs is missing.
These two problems are not necessarily (directly) related, but progress
in one may have implications for the other.

There are good reasons for expecting most complex Hadamard matrices
of order 6 to be elements in a four-parameter set \cite{Bengtsson_08,Skinner_08},
but up till now only one- and two-parameter subsets have been described
on closed form. Recent progress includes the identification of three
new two-parameter families \cite{Szollosi,Karlsson_JMP} that, together
with the two Fourier families, incorporate all previously described
one-parameter families as subfamilies. These five two-parameter families
are partially overlapping, indicating that they might have some unidentified
common feature relevant for a more comprehensive characterization.
A clue to what this feature might be was found in \cite{Karlsson_JMP}
where it was observed that the matrices of the discovered two-parameter
family, $K_{6}^{(2)}$ in the notation of \cite{Tadej webguide},
could be seen as composed of nine $2\times2$ Hadamard submatrices. 

In the present paper it is shown that the set of Hadamard matrices
having such a substructure includes not only $K_{6}^{(2)}$ but also
all other so far described one- and two-parameter Hadamard families
(disregarding families for which there only exists numerical evidence).
More generally, it is shown that any complex Hadamard matrix of order
6 is equivalent to a matrix where either all or none of the nine $2\times2$
submatrices are Hadamard; this is the main result of the present paper.
In a separate paper \cite{Karlsson_3param} it will be shown how the
subset of $H_{2}$-reducible matrices can be fully described on closed
form as a three-parameter Hadamard family.

\section{Preliminaries}

The Hadamard matrices of interest here differ from the more common
ones in that the elements are not restricted to $1$ or $-1$ but
can be any complex number on the unit circle.
\begin{defn}
A square matrix $H$ with complex elements $h_{ij}$ is Hada\-mard
if $|h_{ij}|=1$, and if\begin{equation}
HH^{\dagger}=H^{\dagger}H=NE.\label{eq:unitarity}\end{equation}
Here, $N$ is the order of $H$, and $E$ is the unit matrix of order
$N$. 
\end{defn}
The condition (\ref{eq:unitarity}) will be referred to as the unitarity
constraint on $H$, with the understanding that it is the matrix $H/\sqrt{N}$
that is unitary. Furthermore, $HH^{\dagger}=NE$ implies $H^{\dagger}H=NE$,
and vice versa.
\begin{defn}
Two Hadamard matrices are termed equivalent, $H_{1}\sim H_{2}$, if
they can be related through\begin{equation}
H_{2}=D_{2}P_{2}H_{1}P_{1}D_{1}\end{equation}
where $D_{1}$ and $D_{2}$ are diagonal, unitary matrices and $P_{1}$
and $P_{2}$ are permutation matrices. 
\end{defn}
A set of equivalent Hadamard matrices can be represented by a dephased
matrix, with all elements in the first row and the first column equal
to $1$. For order 2, all Hadamard matrices are equivalent to the
dephased matrix\begin{equation}
F_{2}=\left(\begin{array}{cc}
1 & 1\\
1 & -1\end{array}\right)\label{eq:F2}\end{equation}
For orders 3, 4 and 5, all inequivalent complex Hadamard matrices
have been fully characterized, while for order 6  the characterization
is far from complete. Currently it is based on an isolated matrix
$S_{6}^{(0)}$, on the two-parameter (Fourier) families $F_{6}^{(2)}$
and $(F_{6}^{(2)})^{T}$, and on the three recently reported two-parameter
families $K_{6}^{(2)}$, $X_{6}^{(2)}$ and $(X_{6}^{(2)})^{T}$ (all
in the notation of \cite{Tadej guide,Tadej webguide}). 

As a step in the search for a more comprehensive characterization,
the following subset of Hadamard matrices is identified.
\begin{defn}
A complex Hadamard matrix of order 6 is $H_{2}$-reducible if it is
equivalent to a Hadamard matrix for which all the nine $2\times2$
submatrices are Hadamard. 
\end{defn}
The introduction and investigation of $H_{2}$-reducible Hadamard
matrices has turned out to be rewarding, as is detailed in the next
section and in \cite{Karlsson_3param}.

\section{$H_{2}$-reducible Hadamard matrices}

$H_{2}$-reducible Hadamard matrices are more prevalent than might
be thought. The general nature of these matrices is made clear by
the following theorem, which also contains the main result of the
present paper.
\begin{thm}
\label{thm:Let-H-be}Let $H$ be a Hadamard matrix of order 6, with
elements $h_{ij}$, $i,j=1,$...6. If there exists an order 2 submatrix
$\left(\begin{array}{cc}
h_{ij} & h_{ik}\\
h_{lj} & h_{lk}\end{array}\right)$ that is Hadamard, then $H$ is $H_{2}$-reducible.
\end{thm}
As a corollary it will be seen that all currently known one- and two-parameter
Hadamard families are equivalent to subsets in the set of $H_{2}$-reducible
Hadamard matrices; in contrast, the isolated matrix $S_{6}^{(0)}$
turns out not to be $H_{2}$-reducible.

\textcompwordmark{}

The proof of Theorem \ref{thm:Let-H-be} proceeds in several steps.
First recall the following properties of the elements of Hadamard
matrices.
\begin{lem}
\label{lem:rhomb}Let $z_{1},...,z_{4}$ be four complex numbers on
the unit circle. If $z_{1}+z_{2}+z_{3}+z_{4}=0$, then for each $z_{i}$
there is a $z_{j}$ such that $z_{i}+z_{j}=0$. 
\end{lem}
The proof is immediate since the relation $z_{1}+z_{2}+z_{3}+z_{4}=0$
corresponds to a (possibly degenerate) rhomb in the complex plane.
\begin{lem}
\label{lem:w=00003D0}Let $z_{1}$ and $z_{2}$ be two complex numbers
on the unit circle such that $z_{2}\ne\pm z_{1}$. If $\,\,\mathcal{R}e(z_{1}w)=\mathcal{R}e(z_{2}w)=0$
for some complex number $w$, then $w=0$.
\end{lem}
Again, the proof is elementary.
\begin{prop}
\label{pro:Stand_form}Let $H$ be a Hadamard matrix of order 6 with
elements $h_{ij}$, $i,j=1,$...6. If there exists an order 2 submatrix
$\left(\begin{array}{cc}
h_{ij} & h_{ik}\\
h_{lj} & h_{lk}\end{array}\right)$ that is Hadamard, then $H$ is equivalent to a dephased Hadamard
matrix on the form\begin{equation}
\left(\begin{array}{cccccc}
1 & \,\,\,1 & \,\,\,1 & \,\,\,\,1 & \,\,\,\,1 & \,\,\,\,1\\
1 & -1 & \,\,\,\, z_{1} & \,-z_{1} & \,\,\,\, z_{2} & \,-z_{2}\\
1 & \,\,\, z_{3} & \bullet & \bullet & \bullet & \bullet\\
1 & -z_{3} & \bullet & \bullet & \bullet & \bullet\\
1 & \,\,\, z_{4} & \bullet & \bullet & \bullet & \bullet\\
1 & -z_{4} & \bullet & \bullet & \bullet & \bullet\end{array}\right)\label{eq:Standard form}\end{equation}
\end{prop}
\begin{proof}
Through permutation of rows and columns, the submatrix \linebreak{}
 $\left(\begin{array}{cc}
h_{ij} & h_{ik}\\
h_{lj} & h_{lk}\end{array}\right)$ can be brought to the upper left corner of $H$. A subsequent dephasing
turns it into $F_{2}=\left(\begin{array}{cc}
1 & 1\\
1 & -1\end{array}\right)$, and an overall dephasing results in a matrix on the form\[
\left(\begin{array}{cccccc}
1 & \,\,\,1 & \,\,1 & \,\,\,1 & \,\,\,1 & \,\,\,1\\
1 & -1 & \: u_{1} & \, u_{2} & \: u_{3} & \: u_{4}\\
1 & \,\, w_{1} & \bullet & \bullet & \bullet & \bullet\\
1 & \,\, w_{2} & \bullet & \bullet & \bullet & \bullet\\
1 & \,\, w_{3} & \bullet & \bullet & \bullet & \bullet\\
1 & \,\, w_{4} & \bullet & \bullet & \bullet & \bullet\end{array}\right)\]
where all $u_{i}$ and $w_{i}$ are on the unit circle. The unitarity
constraint now requires that $u_{1}+u_{2}+u_{3}+u_{4}=0$, and that
$w_{1}+w_{2}+w_{3}+w_{4}=0$. These relations can only be satisfied
if to each $u_{i}$ there is a $u_{k}=-u_{i}$, and similarly for
$w_{i}$ (Lemma \ref{lem:rhomb}). A final permutation of rows and
of columns, and a renaming of the entries, leaves the matrix on the
standard form (\ref{eq:Standard form}). 
\end{proof}
At this point it is convenient to introduce the four Hadamard matrices\begin{equation}
\begin{array}{ccccccc}
Z_{1} & = & \left(\begin{array}{cc}
1 & 1\\
z_{1} & -z_{1}\end{array}\right) & \hspace{10ex} & Z_{2} & = & \left(\begin{array}{cc}
1 & 1\\
z_{2} & -z_{2}\end{array}\right)\\
\\Z_{3} & = & \left(\begin{array}{cc}
1 & z_{3}\\
1 & -z_{3}\end{array}\right) &  & Z_{4} & = & \left(\begin{array}{cc}
1 & z_{4}\\
1 & -z_{4}\end{array}\right)\end{array}\label{eq:Z1Z2Z3Z4}\end{equation}
and write the matrix (\ref{eq:Standard form}) on block form\begin{equation}
H=\left(\begin{array}{ccc}
F_{2} & Z_{1} & Z_{2}\\
Z_{3} & a & b\\
Z_{4} & c & d\end{array}\right)\label{eq:Short standard form}\end{equation}
The remaining task is to show that any Hadamard matrix of the type
specified in Theorem \ref{thm:Let-H-be} is equivalent to (or equals)
a matrix on the form (\ref{eq:Short standard form}) where also the
four $2\times2$ submatrices $a$, $b$, $c$, and $d$ are Hadamard.
\begin{prop}
\label{pro:one gives all}If a Hadamard matrix has the form (\ref{eq:Short standard form}),
and one of the matrices $a$, $b$, $c$ and $d$ is Hadamard, then
the other three are also Hadamard.\end{prop}
\begin{proof}
The unitarity constraints (\ref{eq:unitarity}) imply, among other
relations, that\begin{eqnarray}
aa^{\dagger}+bb^{\dagger} & = & 4e\label{eq:Constr aa!+bb!}\\
cc^{\dagger}+dd^{\dagger} & = & 4e\label{eq:Constr cc!+dd!}\\
a^{\dagger}a+c^{\dagger}c & = & 4e\label{eq:Constr a!a+c!c}\\
b^{\dagger}b+d^{\dagger}d & = & 4e\label{eq:Constr b!b+d!d}\end{eqnarray}
Let $a$ be Hadamard. Then the relations (\ref{eq:Constr aa!+bb!})
and (\ref{eq:Constr a!a+c!c}) reduce to $bb^{\dagger}=2e$ and $c^{\dagger}c=2e$,
i.e $b$ and $c$ are also Hadamard. It now follows from (\ref{eq:Constr cc!+dd!})
that $dd^{\dagger}=2e$, i.e. also $d$ is Hadamard. Similar arguments
apply if $b$, $c$ or $d$ is chosen as the initially Hadamard matrix.
\end{proof}
For completeness, the following result from \cite{Karlsson_JMP} is
included here. This result initiated the present investigation.
\begin{thm}
\label{thm:K_6 case}If a Hadamard matrix has the form (\ref{eq:Short standard form}),
and $Z_{1}=Z_{2}$ and $Z_{3}=Z_{4}$, then $H$ is equivalent to
a Hadamard matrix on the same form where $a$, $b$, $c$ and $d$
are Hadamard, with a=d and b=c.\end{thm}
\begin{proof}
The unitarity constraints (\ref{eq:unitarity}) give rise to four
linear relations between $a$, $b$, $c$ and $d$,\begin{equation}
a+b=a+c=b+d=c+d=-Z\label{eq:uni_sum}\end{equation}
where $Z=Z_{3}F_{2}Z_{1}/2$. These relation imply that $d=a$ and
$c=b$, and that the remaining unitarity constraints can be simplified,
to read\begin{equation}
(a-b)^{\dagger}(a-b)=(a-b)(a-b)^{\dagger}=6e.\label{eq:uni-diff}\end{equation}
The matrix $Z$ has the property that $Z^{\dagger}Z=ZZ^{\dagger}=2e$,
and the matrix elements satisfy the relations $Z_{21}=z_{1}z_{2}\bar{Z}_{12}$,
$Z_{22}=-z_{1}z_{2}\bar{Z}_{11}$ and $|Z_{ij}|^{2}\le2$. 

Since the modulus of each element of $a$ and $b$ is one, the relation
$a+b=-Z$ can be solved element by element,\begin{eqnarray*}
a_{ij} & = & -Z_{ij}(\frac{1}{2}+i\sigma_{ij}\sqrt{\frac{1}{|Z_{ij}|^{2}}-\frac{1}{4}})\\
b_{ij} & = & -Z_{ij}(\frac{1}{2}-i\sigma_{ij}\sqrt{\frac{1}{|Z_{ij}|^{2}}-\frac{1}{4}})\end{eqnarray*}
where $\sigma_{ij}=\pm1$. The relations (\ref{eq:uni-diff}) simply
impose further constraints on the sign factors $\sigma_{ij}$,\begin{equation}
\sigma_{11}\sigma_{21}=\sigma_{12}\sigma_{22}.\label{eq:sigma_constraint}\end{equation}
Through permutation of the rows and/or the columns of $H$, it can
be verified that all sign combinations compatible with (\ref{eq:sigma_constraint})
correspond to equivalent matrices. If in particular the sign factors
are related through $\sigma_{11}+\sigma_{22}=\sigma_{12}+\sigma_{21}=0$,
then $a^{\dagger}a=b^{\dagger}b=2e$. Therefore, $H$ is equivalent
to a matrix for which all the $2\times2$ submatrices are Hadamard,
as was to be shown. 
\end{proof}
Theorem \ref{thm:Let-H-be} can now be proven.
\begin{proof}
In (\ref{eq:Short standard form}), let $a=\frac{1}{2}Z_{3}AZ_{1}$,
$b=\frac{1}{2}Z_{3}BZ_{2}$, $c=\frac{1}{2}Z_{4}CZ_{1}$ and $d=\frac{1}{2}Z_{4}DZ_{2}$.
The unitarity constraints (\ref{eq:unitarity}) on $H$ give rise
to four linear relations between $A$, $B$, $C$ and $D$, \begin{equation}
\left\{ \begin{array}{ccc}
A & + & B=-F_{2}\\
C & + & D=-F_{2}\\
A & + & C=-F_{2}\\
B & + & D=-F_{2}\end{array}\right.\label{eq:LinearUnit}\end{equation}
and these relations imply that $D=A$ and $C=B$. As a result,\begin{eqnarray*}
a & = & \left(\begin{array}{cc}
a_{11}(z_{3},z_{1}) & a_{11}(z_{3},-z_{1})\\
a_{11}(-z_{3},z_{1}) & a_{11}(-z_{3},-z_{1})\end{array}\right)\\
b & = & \left(\begin{array}{cc}
b_{11}(z_{3},z_{2}) & b_{11}(z_{3},-z_{2})\\
b_{11}(-z_{3},z_{2}) & b_{11}(-z_{3},-z_{2})\end{array}\right)\\
c & = & \left(\begin{array}{cc}
b_{11}(z_{4},z_{1}) & b_{11}(z_{4},-z_{1})\\
b_{11}(-z_{4},z_{1}) & b_{11}(-z_{4},-z_{1})\end{array}\right)\\
d & = & \left(\begin{array}{cc}
a_{11}(z_{4},z_{2}) & a_{11}(z_{4},-z_{2})\\
a_{11}(-z_{4},z_{2}) & a_{11}(-z_{4},-z_{2})\end{array}\right)\end{eqnarray*}
where\begin{eqnarray*}
a_{11}(z_{3},z_{1}) & = & (A_{11}+z_{1}A_{12}+z_{3}A_{21}+z_{1}z_{3}A_{22})/2\\
b_{11}(z_{3},z_{2}) & = & (B_{11}+z_{2}B_{12}+z_{3}B_{21}+z_{2}z_{3}B_{22})/2\end{eqnarray*}
If it can be shown that $A$ satisfies the unitarity constraint $A^{\dagger}A=2e$,
then $a^{\dagger}a=2e$, $a$ is Hadamard, and a reference to Propositions
\ref{pro:Stand_form} and \ref{pro:one gives all} completes the proof. 

The elements of $A$ are constrained by the condition that all elements
of $a$, $b$, $c$ and $d$ are on the unit circle, and this condition
is sufficient to ensure that $A^{\dagger}A=2e$. Indeed, from the
conditions $|a_{ij}|=1$ one finds\begin{eqnarray}
|A_{11}|^{2}+|A_{12}|^{2}+|A_{21}|^{2}+|A_{22}|^{2} & = & 4\label{eq:A_cond_1}\\
\mathcal{R}e(z_{3}(A_{21}\bar{A}_{11}+A_{22}\bar{A}_{12})) & = & 0\label{eq:A_cond_2}\\
\mathcal{R}e(z_{1}(A_{12}\bar{A}_{11}+A_{22}\bar{A}_{21})) & = & 0\label{eq:A_cond_3}\\
\mathcal{R}e(z_{1}z_{3}A_{22}\bar{A}_{11}+\frac{z_{1}}{z_{3}}A_{12}\bar{A}_{21})) & = & 0\label{eq:A_cond_4}\end{eqnarray}
The conditions on the elements of $d$ give rise to a similar set
of equations, with $z_{1}\to z_{2}$ and $z_{3}\to z_{4}$. From the
elements of $b$ and $c$ there are two more sets, which are obtained
from (\ref{eq:A_cond_1})-(\ref{eq:A_cond_4}) by taking $A\to B$,
and $z_{1}\to z_{2}$ (for $b$), or $z_{3}\to z_{4}$ (for $c$).
The last two sets can be converted into conditions on the elements
of $A$ by means of the relation $B=-F_{2}-A$. The resulting set
of equations can be simplified using the relations (\ref{eq:A_cond_1})-(\ref{eq:A_cond_4}),
and read \begin{eqnarray}
\mathcal{R}e(A_{11}+A_{12}+A_{21}-A_{22}) & = & -2\label{eq:A_cond_5}\\
\mathcal{R}e(z_{3}(\bar{A}_{11}+A_{22}+A_{21}-\bar{A}_{12})) & = & 0\label{eq:A_cond_6}\\
\mathcal{R}e(z_{2}(\bar{A}_{11}+A_{22}+A_{12}-\bar{A}_{21})) & = & 0\label{eq:A_cond_7}\\
\mathcal{R}e(z_{2}z_{3}(A_{22}-1)(\bar{A}_{11}+1)+\frac{z_{2}}{z_{3}}(A_{12}+1)(\bar{A}_{21}+1)) & = & 0,\label{eq:A_cond_8}\end{eqnarray}
from $b$, and there is a similar set, with $z_{2}\to z_{1}$ and
$z_{3}\to z_{4}$, from $c$. Several cases need to be distinguished.

\textcompwordmark{}

\emph{Case 1}. $z_{1}\ne\pm z_{2}$ and $z_{3}\ne\pm z_{4}$. 

From (\ref{eq:A_cond_2}) and the corresponding equation with $z_{3}\to z_{4}$
it follows from Lemma \ref{lem:w=00003D0} that\begin{equation}
A_{21}\bar{A}_{11}+A_{22}\bar{A}_{12}=0.\label{eq:A_cond_9}\end{equation}
Similarly, from (\ref{eq:A_cond_3}) and the corresponding equation
with $z_{1}\to z_{2}$ it follows that\begin{equation}
A_{12}\bar{A}_{11}+A_{22}\bar{A}_{21}=0.\label{eq:A_cond_10}\end{equation}
As a result, $|A_{12}|=|A_{21}|$ and $|A_{11}|=|A_{22}|$, and, from
(\ref{eq:A_cond_1}), $|A_{11}|^{2}+|A_{21}|^{2}=2$. The matrix $A$
therefore satisfies the unitarity constraint $A^{\dagger}A=2e$, as
was to be shown.

\textcompwordmark{}

\emph{Case 2}. $z_{1}\ne\pm z_{2}$ but $z_{3}=\pm z_{4}$. 

From (\ref{eq:A_cond_3}), (\ref{eq:A_cond_4}), (\ref{eq:A_cond_7})
and (\ref{eq:A_cond_8}), and the corresponding relations where $z_{1}\leftrightarrow z_{2}$,
it follows that (since $z_{3}=\pm z_{4}$ and by Lemma \ref{lem:w=00003D0})\begin{eqnarray}
A_{12}\bar{A}_{11}+A_{22}\bar{A}_{21} & = & 0\label{eq:A_cond_11}\\
z_{3}A_{22}\bar{A}_{11}+\frac{1}{z_{3}}A_{12}\bar{A}_{21} & = & 0\label{eq:A_cond_12}\\
\bar{A}_{11}+A_{22}+A_{12}-\bar{A}_{21} & = & 0\label{eq:A_cond_13}\\
z_{3}(A_{22}-1)(\bar{A}_{11}+1)+\frac{1}{z_{3}}(A_{12}+1)(\bar{A}_{21}+1) & = & 0\label{eq:A_cond_14}\end{eqnarray}
Combining (\ref{eq:A_cond_11}) and (\ref{eq:A_cond_13}), and (\ref{eq:A_cond_12})
and (\ref{eq:A_cond_14}), one finds the conditions\begin{eqnarray}
(\bar{A}_{11}-\bar{A}_{21})(A_{12}-\bar{A}_{21}) & = & 0\label{eq:A_cond_15}\\
z_{3}(A_{22}-\bar{A}_{11}-1)+\frac{1}{z_{3}}(A_{12}+\bar{A}_{21}+1) & = & 0\label{eq:A_cond_16}\end{eqnarray}
In view of (\ref{eq:A_cond_15}), either $A_{21}=\bar{A}_{12}$ or
$A_{21}=A_{11}$.

\textcompwordmark{}

\emph{Subcase 2.1.} Let $A_{21}=\bar{A}_{12}$. By (\ref{eq:A_cond_13}),
$A_{22}=-\bar{A}_{11}$, and hence, by (\ref{eq:A_cond_1}), $|A_{11}|^{2}+|A_{21}|^{2}=|A_{12}|^{2}+|A_{22}|^{2}=2$.
This relation, together with (\ref{eq:A_cond_11}), implies that $A^{\dagger}A=2e$. 

\textcompwordmark{}

\emph{Subcase 2.2.} Let instead $A_{21}=A_{11}$. Then, by (\ref{eq:A_cond_13}),
$A_{12}=-A_{22}$, and hence, by (\ref{eq:A_cond_6}), (\ref{eq:A_cond_12})
and (\ref{eq:A_cond_16}),\begin{eqnarray*}
(|A_{11}|^{2}-|A_{22}|^{2})\mathcal{R}e(z_{3}) & = & 0\\
A_{22}\bar{A}_{11}\mathcal{I}m(z_{3}) & = & 0\\
(1+\bar{A}_{11}-A_{22})\mathcal{I}m(z_{3}) & = & 0\end{eqnarray*}
If here $\mathcal{R}e(z_{3})\ne0$, then $|A_{11}|^{2}=|A_{22}|^{2}=|A_{12}|^{2}=|A_{21}|^{2}=1$
and again $A^{\dagger}A=2e$ (with the additional condition that $\mathcal{I}m(z_{3})=0$,
i.e. $z_{3}^{2}=z_{4}^{2}=1$). 

If instead $\mathcal{R}e(z_{3})=0$, so that $\mathcal{I}m(z_{3})\ne0$,
then either $A_{11}=A_{21}=0$ with $A_{22}=-A_{12}=1$, or $A_{11}=A_{21}=-1$
with $A_{22}=-A_{12}=0$. Neither of these conditions is compatible
with the condition (\ref{eq:A_cond_1}), expressing that there exists
no Hadamard matrix on the form (\ref{eq:Short standard form}) such
that $\mathcal{R}e(z_{3})=\mathcal{R}e(z_{4})=0.$ 

Summarizing Case 2, for $H$ to be Hadamard, either $A_{12}=\bar{A}_{21}$
and $A_{22}=-\bar{A}_{22}$, or else $A_{11}=A_{21}$ and $A_{12}=-A_{22}$,
with the additional condition $z_{3}^{2}=z_{4}^{2}=1$. In either
case, $A$ satisfies the unitarity constraint $A^{\dagger}A=2e$,
as was to be shown.

\textcompwordmark{}

\emph{Case 3}. $z_{1}=\pm z_{2}$ but $z_{3}\ne\pm z_{4}$. 

The arguments for this case mirror those of Case 2.

\textcompwordmark{}

\emph{Case 4}. $z_{1}=\pm z_{2}$ and $z_{3}=\pm z_{4}$. 

This case is covered by Theorem \ref{thm:K_6 case}. 

\textcompwordmark{}

In all cases, the matrix $A$ therefore satisfies the unitarity constraint
$A^{\dagger}A=2e$, and with this result, the proof of Theorem \ref{thm:Let-H-be}
is completed.

\textcompwordmark{}
\end{proof}
From its dephased form, it is easy to see whether a Hadamard matrix
is $H_{2}$-reducible or not.
\begin{cor}
Let H be a complex Hadamard matrix of order 6. H is $H_{2}$-reducible
if, and only if, its dephased form has at least one element equal
to -1. \end{cor}
\begin{proof}
If one element equals $-1$, there is a submatrix which equals the
$2\times2$ Hadamard matrix $F_{2}=\left(\begin{array}{cc}
1 & 1\\
1 & -1\end{array}\right)$, and Theorem \ref{thm:Let-H-be} applies. On the other hand, if a
dephased Hadamard is reducible, the upper left corner $2\times2$
Hadamard submatrix must equal $F_{2}$.
\end{proof}
It follows from the corollary that all the currently known \cite{Tadej webguide}
one- and two-parameter families of order 6 are families of $H_{2}$-reducible
Hadamard matrices. On the other hand, the single, isolated matrix
$S_{6}^{(0)}$ is not $H_{2}$-reducible.

\section{MUBs and $H_{2}$-reducible Hadamard matrices}

As was pointed out above, all known, closed form Hadamard families
of order 6 are families of $H_{2}$-reducible Hadamard matrices. A
similar statement holds for the few cases where closed form MUB matrices
are known. For instance, let $\{I,\, F_{6}(0,b),\, C(b)\}$ be the
family of MUB triplets as presented in Theorem 2.4 of \cite{Jaming}.
As is easily verified, for each $b$, $C(b)$ is equivalent to $F_{6}(0,b')$
for some $b'$, and $C(b)$, like $F_{6}(0,b)$, therefore belongs
to the set of $H_{2}$-reducible Hadamard matrices. Similarly, Zauner's
construction \cite{Zauner}, as quoted in \cite{Jaming}, involves
a family of triplets $\{I,\, E_{1}(x),\, E_{2}(x)\}$. For each $x$,
the matrices $E_{1}$ and $E_{2}$ are both equivalent to $F_{6}(0,0)$
(from  (B.1-2) in \cite{Jaming}), and, as noted in \cite{Jaming},
$E_{1}^{\dagger}E_{2}$ is equivalent to a member of the family $D_{6}^{(1)}$
(in the notation of \cite{Tadej guide,Tadej webguide}). Again, therefore,
$E_{1}$, $E_{2}$ and $E_{1}^{\dagger}E_{2}$ are all in the set
of $H_{2}$-reducible Hadamard matrices.

\section{Conclusion and outlook}

In a separate paper \cite{Karlsson_3param} it is shown that an $H_{2}$-reducible
Hadamard matrix can be fully characterized in terms of a three-parameter
family of complex Hadamard matrices of order 6. The overall picture
is therefore that the subset of $H_{2}$-reducible Hadamard matrices
has been completely characterized, and that in the process all previously
described one- and two-parameter families reappear in a unified setting. 

For the set of Hadamard matrices that are not $H_{2}$-reducible,
on the other hand, very little is known: it contains the isolated
matrix $S_{6}^{(0)}$, and some of its members belong to one or several
four-parameter families. In spite of recent efforts towards finding
Hadamard families, not a single (analytically described) family has
been found that extends into the set of non-reducible Hadamard matrices.
The additional information that has come from numerical investigations
is also very limited. As expected, numerically generated Hadamard
matrices are in general not $H_{2}$-reducible, unless specifically
designed to be so. Such matrices can also be designed to trace out
subfamilies in the non-reducible domain (from observations of some
$10^{5}$ matrices generated in a semi-random manner; see also \cite{Skinner_08}),
but this is also as expected if indeed a four-parameter family exists.
In all, however, the notion of $H_{2}$-reducibility provides a new
perspective also in the search for a characterization of the full
set of complex Hadamard matrices of order 6.

The concept of $H_{2}$-reducible Hadamard matrices has in this paper
only been defined for order 6. It would seem worthwhile to generalize
this concept to higher orders, by distinguishing the Hadamard matrices
with a substructure of Hadamard blocks from those for which such a
structure is absent. Based on the experience for order 6, the real
challenge will most likely be to find and characterize the Hadamard
matrices that lack such substructure.

The possible relevance of the result obtained here for the understanding
of mutually unbiased bases (MUBs) in six dimensions is left for further
study. Extensive numerical searches \cite{Brierley_08,Brierley_09}
indicate that the maximal number of such bases is no greater than
three, but an understanding of why this should be so is lacking. Similarly,
MUBs, like the Hadamard matrices, come in families \cite{Jaming,Szollosi,Brierley+W+B},
but a full characterization of for instance all triplets of MUBs in
six dimensions has so far not been achieved. Interestingly, all currently
known (to us) Hadamard members of MUB triplets are $H_{2}$-reducible,
even those obtained through numerical searches, and if this observation
reflects a general feature of the MUBs, it may contribute to the understanding
of why no larger sets of MUBs are found in six dimensions.

\end{document}